\documentclass[
preprint,   
 amsmath,amssymb, 
 aps,
]{revtex4-1}
\usepackage{graphicx}
\usepackage{dcolumn}
\usepackage{bm}
\usepackage{soul,xcolor}
\usepackage{CJK}
\usepackage{graphics}
\usepackage{tikz}
\usetikzlibrary{decorations.pathmorphing}
\usetikzlibrary{decorations.markings}
\usetikzlibrary{decorations.pathmorphing}
\usetikzlibrary{decorations.pathreplacing}
\usetikzlibrary{math}
\usetikzlibrary{patterns}
\usetikzlibrary{patterns.meta}
\usepgfmodule{oo}
\usepackage{caption}
\usepackage{ragged2e}
\usepgfmodule{oo}

\begin{document}


\title{Super-Enhanced Absorption of Gravitons in  Atomic Gases}

\author{Yongle Yu }
\email{yongle.yu@wipm.ac.cn}

 \affiliation{ 
  Wuhan Institute of Physics  and Mathematics, Chinese Academy of Science,\\
 West No. 30 Xiao Hong Shan, Wuchang, Wuhan, 430071, China \vspace{2em}}

\begin{abstract}
We present a novel method for detecting gravitons using 
an atomic gas supported by laser fields.
Despite the coupling strength of gravitons to atomic 
transitions being orders of magnitude weaker than that of 
photons to atomic transitions, the rate of graviton-absorbed 
atomic transitions can be substantially elevated to a practically 
observable level. This enhancement is facilitated by an 
exceptionally potent amplification effect, stemming from
 a collective quantum electrodynamics phenomenon that 
 encompasses a simultaneous multiphoton-multiatom process.

\end{abstract}
\maketitle

Relativity and quantum theories emerged in the early 
twentieth century. Nearly a century later, the direct observation 
of gravitational waves (GWs) validated a fundamental prediction of 
general relativity \cite{abbott2016observation, abbott2016gw151226, abbott2017gw170817}. This 
milestone not only ushered in a new 
era of astronomical observation  but 
also significantly propelled efforts to unify quantum theory 
and general relativity. A pivotal question regarding GWs is whether 
these waves are composed of quantum particles known as gravitons. 
Experimental verification of gravitons is essential for 
the unification of quantum and relativity theories.
 Several theoretical studies have investigated
  the direct observation of gravitons 
  through ultraweak atomic 
 transitions involving these particles 
\cite{dyson2004world, dyson2013graviton,Rothman2006can, BoughnHydrogenGraviton2006,Pitelli2021angular,Hu2021high,
 carney2024grvitonclassicalExploration}. Some proposals
 for testing the quantum nature of gravity, without 
 detecting gravitons directly, focus on the signals 
 arising from gravity's quantum properties. These 
 include gravity-induced entanglement between two masses  
 \cite{Marletto2017TwoMass, SpinEntanglement}, 
  and the fluctuation of the arm length
  in a GW detector due to quantum 
 states of GWs \cite{Wilczek1, Wilczek2, Zhangbaocheng, Kanno2021}.
 A recent study \cite{tobar2024detecting} by Tobar {\it{et al.}} introduces a 
 novel approach for detecting single gravitons
  in the low-frequency range, highlighting a gravito-phononic analog of 
 the photo-electric effect enabled by advancements in quantum 
 resonators and continuous measurement techniques.
 In this paper, we propose that ultraweak graviton-absorption atomic
 transitions can be amplified to observable levels using a recently 
 uncovered quantum enhancement mechanism. This makes the direct 
 observation of gravitons  readily realizable
 in a laboratory setting without necessitating further 
 technological advancements.

The detection of gravitons could, in principle, be achieved 
through graviton-mediated atomic transitions, analogous to photon-mediated 
atomic transitions, where, in a simplified approximation,  
the electron mass acts as the gravitational
charge coupling to the quantum gravitational field. 
Given the extremely weak gravitational coupling—approximately $10^{-43}$ times 
weaker than the electromagnetic coupling \cite{Gcoupling}—it appears challenging 
to avoid the conclusion that observing gravitons via atomic transitions remains 
impractical in the near future.
However, our recent study  shows that an ultraweak atomic transition 
can be significantly amplified by integrating it with a multiphoton-multiatom 
(MPMA) process \cite{yu2}. This amplification reveals several 
noteworthy characteristics: it not only delivers substantial enhancements 
to the transition rate, potentially elevating it by several tens of 
orders of magnitude, but also exhibits a near-saturation behavior.

In \cite{yu2}, we explored certain ultraweak atomic transitions, 
such as higher electric multipole $Ej$ transitions ($j=3,4, \dots$), in 
atoms mediated by the absorption of a corresponding $Ej$ photon,
which  can be amplified to observable levels through the MPMA process. 
 A general analysis of the strength of these atomic transitions, 
 alongside a comparison with graviton-mediated atomic transitions, offers 
 valuable insights. 
The  $Ej$ transition probability (where  $j=1,2$ correspond 
to electric dipole and quadrupole transitions, respectively) scales approximately 
as $(a/\lambda)^{2j}$, where $a$ denotes the linear size of the atom and 
$\lambda$ represents the wavelength of the involved photon \cite{berestetskiiQEDbook}. 
Typically, $(a/\lambda) \approx 10^{-4}$. Consequently, the $E2$ transition probability 
is on the order of $10^{-8}$ relative to the $E1$ transition, the $E3$ transition 
on the order of $10^{-16}$, $\dots$, the $E7$ transition on the order 
of $10^{-48}$, and the $E8$ transition on the order of $10^{-56}$.
The graviton-mediated transition
is anticipated to be on the order of $10^{-(8+43)} = 10^{-51}$, where the 
factor $10^{-8}$ arises due to that the graviton field is a second-rank 
tensor, similar to an $E2$ field. For instance, in a hydrogen atom, the $E1$ decay
rate (the $2p-1s$ transition) is on the order of $10^9$ s$^{-1}$. 
Hence, the graviton-mediated transition rate  is projected to be on the 
order of $10^{-51} \times 10^9$ $= 10^{-42}$ s$^{-1}$,  
compared to a rate of $5.7 \times 10^{-40}$ s$^{-1}$ for the $3d-1s$ transition, as 
calculated in \cite{BoughnHydrogenGraviton2006}. This graviton-mediated transition
exhibits a rate relatively close to that of an $E7$ transition but exceeds
that of an $E8$ transition.
 Quantum enhancement through the MPMA process could potentially 
enable the observation of even higher $Ej$ photoabsorption 
transitions for $j=9,10,\dots$, thereby making it feasible to 
amplify the graviton-absorption transition to an observable level.

The MPMA process in an atomic gas represents a high-order quantum electrodynamics (QED)
 phenomenon wherein a specific number of atoms undergo a cooperative 
 transition by simultaneously absorbing laser photons \cite{WhiteAtomgasTlBa, NaAtoms, 
 rios1980lineshape, andrews1983cooperative, 
 Nayfeh1984, Kim1998, Muthukrishnan2004, Zheng2013, twoMolecules, yu1,yu2}.
 Detailed analyses of the MPMA process and its distinctive properties were 
 presented in our recent studies \cite{yu1, yu2}.
 Here, we 
provide a simplified 
introduction to the process and its enhancement capabilities.

Consider an atomic gas consisting of two species of atoms, designated 
as $A$-species and $B$-species. (The introduction of two distinct 
species is primarily for formal convenience; a single-species scheme can also 
be naturally implemented—see \cite{yu1}.)
 Within this gas, we examine an $m$-atom 
system composed of one $A$-species atom and $m-1$ $B$-species atoms. The 
$A$-species atom can undergo an $E1$ transition, characterized by
 an angular transition frequency $\omega_a$ between the ground 
state $|g_a\rangle$ and an excited state $|e_a\rangle$, while each 
$B$-species atom possesses an $E1$ transition from the ground state 
$|g_b\rangle$ to an excited state $|e_b\rangle$, with a transition frequency $\omega_b$.

Two lasers,  labeled as $\mathfrak{L}_1$ and $\mathfrak{L}_2$ with frequencies $\Omega_{\mathfrak{L}_1}$
 and $\Omega_{\mathfrak{L}_2}$, respectively,  are utilized
to induce  atomic transitions. If $\Omega_{\mathfrak{L}_1}$ is 
tuned near, but not equal to, $\omega_a$, a single-photon absorption process cannot 
effectively excite an $A$-species atom. However, the $m$-atom system can 
undergo a simultaneous joint transition provided the laser frequencies 
fulfill the energy conservation condition:
\begin{equation}
\hbar \Omega_{\mathfrak{L}_1} + (m-1) \hbar \Omega_{\mathfrak{L}_2} = \hbar \omega_a + (m-1) \hbar \omega_b,
\end{equation}
where $\hbar$ represents the reduced Planck constant. In this joint excitation process,
 the $A$-species atom absorbs one photon from the laser $\mathfrak{L}_1$, 
 while each of the $m-1$ $B$-species atoms absorbs one photon from 
 the laser $\mathfrak{L}_2$ (see Fig. \ref{fig_3p3a2L} for an example with $m=3$).
 
 \tikzmath{  \bz= 0.4; \bsz=0.5; \bsf=6;    
\wsf=\bsf+5; 
\wint= 0.12;  
\wx= -3.8; \wy=0.8; \wz= 0.7; 
\wsz=1.5;
\asz=0.07;
\rEx= 0.1; \Esz=1.1; \rEy= -1.4;  \rEz=0.2;\rint=2.2; 
\bEx= -0.1;  \bEy= 1.0; \bEz= -0.1;  \bint= 1.0; \brEyg=0.3; 
\Elabsz= 0.85; 
\atomsize= 0.6;  
\pad= 0.05;
\seglen= 8pt; \amp=1.pt; 
\padint= 0.7pt;
 }

\tikzmath{ \arcR=40; \arcAng=5; \arcH=1.8;  \opacity=0.7; \ballR=0.1cm; 
   \centrX=4.4; \atomShift=0.9; \centrY=-0.9; \bEx=1; \bEy=0.5;\rEx= -1.0; \Elabsz= 0.85;
   \rEy= 0.5; \rint= 1.2;\bint=1.0; \bsz=0.5; \bsf=6; \pad=0.05; \seglen= 8pt; \amp=1.pt;\Esz=1.1; 
\padint= 1.3pt; }

\tikzmath{ \arcR=40; \arcAng=5; \arcH=1.8;  \opacity=0.7; \ballR=0.1cm; 
   \centrX=4.4; \atomShift=0.9; \centrY=-0.9; \bEx=1;\rEx= -0.7;  \Elabsz= 0.85;
   \rEy= -0.03; \bEy=\rEy; \rint= 1.2;\bint=1.0; \bsz=0.5;  \pad=0.05; \seglen= 5pt; \amp=0.6pt;\Esz=0.7; 
\padint= 0.4pt; \wsz=0.6; \bEx= \rEx + \Esz+ 0.1; \bsf=0; \wint=0.02; \wx=0.24; \levelshift=-0.2; \wyinz=0.1;\ypanel=3.0;   \xpanel=5.0;
\photontraindx= 1.; }
\definecolor{LaserPink}{RGB}{241,184,186}
\definecolor{lightgreen}{RGB}{68, 85, 90}
\definecolor{lightwebblue}{RGB} {193,225,236}   
\definecolor{webblue}{RGB} {20, 163, 199} 
\definecolor{shiningsteel}{RGB}{192,192,192}
\definecolor{dodgeblue}{RGB}{30, 144, 255}
\pgfooclass{photonline}{

  \method apply(#1,#2,#3,#4,#5,#6) {
   \draw[decorate,decoration={snake, segment length= \seglen, amplitude=\amp},color=#6](#1, #2)  -- +(#3,0);
  \draw[-stealth,color=#6](#1+#4,#2)  -- +(#5*2,0);
  
  }

 \method apply_dashed(#1,#2,#3,#4,#5,#6) {
 \draw[decorate,dash pattern={on \padint off \padint on \padint off \padint on \padint},
 decoration={snake, segment length= \seglen, amplitude=\amp},color=#6](#1, #2)  -- +(#3,0);
  \draw[-stealth,dash pattern={on \padint off \padint on \padint off \padint on \padint}, color=#6](#1+#4,#2)  -- +(#5*2,0);
  
  }
 }
 
 \pgfooclass{photonlineRatio}{

  \method apply(#1,#2,#3,#4,#5,#6,#7) {
   \draw[decorate,decoration={snake, segment length= #7*\seglen, amplitude=#7*\amp},color=#6](#1, #2)  -- +(#3,0);
  \draw[-stealth,color=#6](#1+#4,#2)  -- +(#5*1.5,0);
  
  }

 }
 \pgfooclass{atomTransition}{
   \method two_levels(#1,#2,#3,#4,#5,#6,#7) {
   \draw[thick ,color=#6](#1, #2)  -- +(#3,0);
   \draw[thick ,color=#7](#1, #2+#4)  -- +(#3,0);
  }
   \method excitation(#1,#2,#3,#4,#5,#6){
    
   \draw[color=#6](#1+#3/2, #2)  circle(#5);
   \draw[dashed, color=#6, -stealth] (#1+#3/2, #2)--+(0,#4);
    }
     
   \method atom_frequency(#1,#2,#3,#4,#5,#6, #7){
       \node[color=#6, font=\tiny] at (#1+#3/2+4.5*\pad, #2+#4/2)  {#7};
       }
   \method photon_frequency(#1,#2,#3,#4,#5,#6,#7){
       \node[color=#6, font=\tiny] at (#1+#3/2+4.5*\pad, #2+#4/2)  {#7};
       } 
       
   \method photon_absorption(#1,#2,#3,#4,#5,#6) {
   \draw[dashed, color=#6] (#1, #2+#4)--+(#3,0);
   
   \draw[decorate,
 decoration={snake, segment length= \seglen, amplitude=\amp},color=#6](#1+#3/2, #2)  -- +(0,#4);
 
    }
}

\pgfooclass{gravitonline}{
   \method apply(#1,#2,#3,#4,#5,#6) {
   \draw[decorate,decoration={coil, aspect=1.2, segment length= \seglen*0.75, amplitude=2*\amp},color=#6](#1, #2)  -- +(#3,0);
   \draw[-stealth,color=#6](#1+#4,#2)  -- +(#5*2,0);
  
  }

 \method apply_dashed(#1,#2,#3,#4,#5,#6) {
 \draw[decorate,dash pattern={on \padint off \padint on \padint off \padint on \padint off \padint},
 decoration={coil, aspect=1.2, segment length= \seglen*0.75, amplitude=2*\amp},color=#6](#1, #2)  -- +(#3,0);
  \draw[-stealth,dash pattern={on \padint off \padint on \padint off \padint on \padint off \padint},color=#6](#1+#4,#2) -- +(#5*2,0);  
  }
 }

\pgfooclass{gravitonlineRatio}{
   \method apply(#1,#2,#3,#4,#5,#6,#7) {
   \draw[decorate,decoration={coil, aspect=1.2, segment length= #7*\seglen*0.75, amplitude=2*#7*\amp},color=#6](#1, #2)  -- +(#3,0);
   \draw[-stealth,color=#6](#1+#4,#2)  -- +(#5*1.5,0);
  
  }
 }
\pgfooclass{laserfield}{
   \method apply(#1,#2,#3,#4,#5,#6,#7,#8,#9) {
   \fill[color=#8,rotate=#7] (#1,#2) rectangle ++(#3,#4);
   \draw[color=#9, rotate=#7] (#1+0.02,#2)-- +(0,#4);
   \draw[color=#9, rotate=#7] (#1+#3-0.02,#2)--+(0,#4);
   \fill[color=#9, rotate=#7](#1,#2-#5)rectangle++(#3,#5);
   \draw[fill=#9, color=#9, rotate=#7] (#1+#3/2-#6, #2)-- +(#6,#6)--+(2*#6, 0)--cycle;
  }   
}

\pgfooclass{detector}{
   \method apply(#1,#2,#3,#4) {  
   \draw[line width=0.1mm, rotate=#4](#1, #2)  -- +(#3,0);
   \draw[line width=0.5mm, rotate=#4] (#1,#2) arc(180:0:#3/2);
   \draw[line width=0.1mm, decorate,decoration={coil, aspect=1.3, segment length= \seglen*0.4, amplitude=2*\amp},
                        rotate=#4](#1+#3/2., #2+#3/2.)  -- +(#3*0.8,#3*0.8);
   
    \begin{scope}  
    \clip [rotate=#4] (#1,#2) arc(180:0:#3/2)-- (#1,#2);
     \begin{scope}
        \shade [ball color=blue, opacity=0.9, rotate=#4] (#1,#2) circle (1);
    \end{scope}
  \end{scope}
  
  }

 }
 
 \pgfooclass{metalwall}{
   \method apply(#1,#2,#3,#4,#5,#6) {  
   \begin{scope}
   \clip (-#1, -#2)-- (-#1,#2)-- (-#1+ #3, #2+#4)-- (-#1 + #3, -#2+ #4)--cycle; 
\end{scope}
   \draw[pattern={Lines[
                  distance=1.2mm,
                  angle=45,
                  line width=0.2mm
                 ]},color=shiningsteel,opacity=0.5](-#1, -#2)-- (-#1-#5,-#2)--(-#1-#5,#2)--(-#1,#2)--cycle;
  \draw[ pattern={Lines[
                  distance=1.2mm,
                  angle=45,
                  line width=0.2mm
                 ]},color=shiningsteel,opacity=0.3](-#1, #2)-- (-#1-#5,#2)--(-#1+ #3 - #5, #2+#4)--(-#1+#3,#2+#4)--cycle;

  }

 }

\pgfoonew \myphotonline=new photonline()
\pgfoonew \myphotonlineRatio=new photonlineRatio()
\pgfoonew \mygravitonline=new gravitonline()
\pgfoonew \mygravitonlineRatio=new gravitonlineRatio()
\pgfoonew \mylaserfield=new laserfield()
\pgfoonew \mydetector=new detector()
\pgfoonew \myatomTransition=new atomTransition()
\pgfoonew \mymetalwall=new metalwall()

\makeatother

\tikzmath{\atomtranx=0.0; \atomtrany= 0.0; \atomtranw=1.6; \atomtranh=1.8; \next= \atomtranw*1.2; \brratio=0.7;  \longline= 2; 
       \large=1.4; \linexinc=0.10;  \lineyinc=0.15; \atAx= \wx+ 1.5* \longline*\wsz; \atgap= 0.7; \atAy= -16*\wint; \atBx= \atAx + 0.4*\atgap;
      \atBy= \atAy + 0.9*\atgap; \atCx= \atAx + 0.7*\atgap; \atCy= \atAy + 0.35*\atgap; \levelgap= 0.8; \atAlevelx= \atAx +0.9*\levelgap;
       \atAlevely= \atAy - 1.1*\levelgap; \atBlevelx= \atBx +0.5*\levelgap;
       \atBlevely= \atBy + 0.6*\levelgap; \atClevelx= \atCx + 1.5*\levelgap;
       \atClevely= \atCy + 0.9*\levelgap;\pad=0.05; \padint= 1.3pt; \rbfac=0.7; \lineshift= 3;}
\colorlet{unexcited-bcol}{green!20!blue!20}
\colorlet{unexcited-rcol}{gray!30!blue!10!}
\colorlet{gray2}{gray!70}
\colorlet{bcol}{green!20!blue!80}
\colorlet{rcol}{magenta!60!}
\definecolor{Laserblue_detune}{RGB}{20, 163, 199} 
\definecolor{redgold}{RGB}{235, 84, 6}

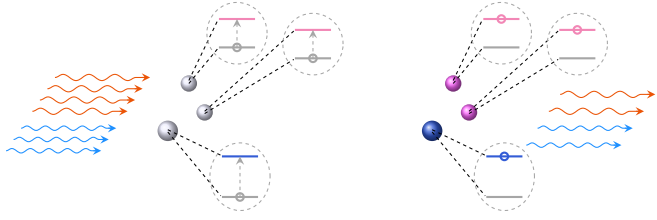
\begin{figure}
\captionsetup{width=.90\linewidth}
\begin{tikzpicture}
\begin{scope}[xshift=0, scale=1.0]

\myphotonlineRatio.apply(\wx -3.5*\linexinc, -3*\wint -3.5*\lineyinc, \longline*\wsz, -2*\wint+\longline*\wsz, \asz, dodgeblue, \large)
\myphotonlineRatio.apply(\wx -2.5*\linexinc, -3*\wint -2.5*\lineyinc, \longline*\wsz, -2*\wint+\longline*\wsz, \asz, dodgeblue, \large)
\myphotonlineRatio.apply(\wx -1.5*\linexinc, -3*\wint -1.5*\lineyinc, \longline*\wsz, -2*\wint+\longline*\wsz, \asz, dodgeblue, \large)
\myphotonlineRatio.apply(\wx, -3*\wint, \longline*\wsz, -2*\wint+ \longline*\wsz, \asz, redgold, 1.4*\large)
\myphotonlineRatio.apply(\wx + \linexinc, -3*\wint + \lineyinc, \longline*\wsz, -2*\wint+\longline*\wsz, \asz, redgold, 1.4*\large)
\myphotonlineRatio.apply(\wx + 2*\linexinc, -3*\wint + 2*\lineyinc, \longline*\wsz, -2*\wint+\longline*\wsz, \asz, redgold, 1.4*\large)
\myphotonlineRatio.apply(\wx + 3*\linexinc, -3*\wint + 3*\lineyinc, \longline*\wsz, -2*\wint+\longline*\wsz, \asz, redgold, 1.4*\large)
\
\node[circle, shading=ball, ball color=unexcited-rcol,  minimum size=2pt, scale=1.25*\atomsize,
text=white] (ball) at(\atAx,  \atAy){} ;
\node[circle, shading=ball, ball color=unexcited-rcol,  minimum size=2pt, scale=1.*\atomsize,
text=white] (ball) at(\atBx,  \atBy){} ;
\node[circle, shading=ball, ball color=unexcited-rcol,  minimum size=2pt, scale=1.*\atomsize,
text=white] (ball) at(\atCx,  \atCy){} ;

\myatomTransition.two_levels(\atAlevelx, \atAlevely, 0.3*\atomtranw, 0.3*\atomtranh, 0, gray2, bcol)
\draw[dash pattern={on \padint off \padint on \padint off \padint on \padint}](\atAx, \atAy-0.3*\pad)--(\atAlevelx-0.3*\pad ,\atAlevely +0.3*\pad );
\draw[dash pattern={on \padint off \padint on \padint off \padint on \padint}](\atAx, \atAy+ 0.3*\pad)--(\atAlevelx-0.3*\pad ,\atAlevely + 0.3*\atomtranh + 0.3*\pad );
\draw[dashed,dash pattern=on \padint off \padint, gray2, -stealth](\atAlevelx+ 0.5*0.3* \atomtranw, \atAlevely )--(\atAlevelx+ 0.5*0.3*\atomtranw, \atAlevely +  0.3*\atomtranh);
\draw[dashed,dash pattern=on \padint off \padint, gray2](\atAlevelx+ 0.5*0.3* \atomtranw, \atAlevely +  0.5*0.3*\atomtranh) ellipse ( 0.25*\atomtranw cm and 0.25*\atomtranh cm);
\draw[thick,gray2](\atAlevelx+ 0.5*0.3* \atomtranw, \atAlevely ) circle (0.05);

\myatomTransition.two_levels(\atBlevelx, \atBlevely, 0.3*\atomtranw, \rbfac * 0.3*\atomtranh, 0, gray2, rcol)
\draw[dash pattern={on \padint off \padint on \padint off \padint on \padint}](\atBx, \atBy-0.0*\pad)--(\atBlevelx-0.3*\pad ,\atBlevely - 0.3*\pad );
\draw[dash pattern={on \padint off \padint on \padint off \padint on \padint}](\atBx, \atBy+ 0.8*\pad)--(\atBlevelx-0.3*\pad ,\atBlevely + 0.3*\rbfac*\atomtranh - 0.3*\pad );
\draw[dashed,dash pattern=on \padint off \padint, gray2, -stealth](\atBlevelx+ 0.5*0.3* \atomtranw, \atBlevely )--(\atBlevelx+ 0.5*0.3*\atomtranw, \atBlevely +  0.3*\rbfac*\atomtranh);
\draw[dashed,dash pattern=on \padint off \padint, gray2](\atBlevelx+ 0.5*0.3* \atomtranw, \atBlevely +  \rbfac*0.5*0.3*\atomtranh) ellipse ( 0.25*\atomtranw cm and 0.25*\atomtranh *\rbfac*1.3 cm);
\draw[thick,gray2](\atBlevelx+ 0.5*0.3* \atomtranw, \atBlevely ) circle (0.05);

\myatomTransition.two_levels(\atClevelx, \atClevely, 0.3*\atomtranw, \rbfac * 0.3*\atomtranh, 0, gray2, rcol)
\draw[dash pattern={on \padint off \padint on \padint off \padint on \padint}](\atCx, \atCy-0.3*\pad)--(\atClevelx-0.3*\pad ,\atClevely - 0.3*\pad );
\draw[dash pattern={on \padint off \padint on \padint off \padint on \padint}](\atCx, \atCy+ 0.3*\pad)--(\atClevelx-0.3*\pad ,\atClevely + 0.3*\rbfac*\atomtranh - 0.3*\pad );
\draw[dashed,dash pattern=on \padint off \padint, gray2, -stealth](\atClevelx+ 0.5*0.3* \atomtranw, \atClevely )--(\atClevelx+ 0.5*0.3*\atomtranw, \atClevely +  0.3*\rbfac*\atomtranh);
\draw[dashed,dash pattern=on \padint off \padint, gray2](\atClevelx+ 0.5*0.3* \atomtranw, \atClevely +  \rbfac*0.5*0.3*\atomtranh) ellipse ( 0.25*\atomtranw cm and 0.25*\atomtranh *\rbfac*1.3 cm);
\draw[thick,gray2](\atClevelx+ 0.5*0.3* \atomtranw, \atClevely ) circle (0.05);

\end{scope}
\begin{scope}[xshift=110, scale=1.0]

\myphotonlineRatio.apply(\wx -3.0*\linexinc+\lineshift, -3*\wint -3.0*\lineyinc, \longline*\wsz, -2*\wint+\longline*\wsz, \asz, dodgeblue, \large)
\myphotonlineRatio.apply(\wx -1.5*\linexinc+\lineshift, -3*\wint -1.5*\lineyinc, \longline*\wsz, -2*\wint+\longline*\wsz, \asz, dodgeblue, \large)
\myphotonlineRatio.apply(\wx +\lineshift, -3*\wint, \longline*\wsz, -2*\wint+ \longline*\wsz, \asz, redgold, 1.4*\large)
\myphotonlineRatio.apply(\wx + 1.5*\linexinc +\lineshift, -3*\wint + 1.5*\lineyinc, \longline*\wsz, -2*\wint+\longline*\wsz, \asz, redgold, 1.4*\large)
\end{scope}

\begin{scope}[xshift=100, scale=1.0]
\node[circle, shading=ball, ball color=bcol,  minimum size=2pt, scale=1.25*\atomsize,
text=white] (ball) at(\atAx,  \atAy){} ;
\node[circle, shading=ball, ball color=rcol,  minimum size=2pt, scale=1.*\atomsize,
text=white] (ball) at(\atBx,  \atBy){} ;
\node[circle, shading=ball, ball color=rcol,  minimum size=2pt, scale=1.*\atomsize,
text=white] (ball) at(\atCx,  \atCy){} ;

\myatomTransition.two_levels(\atAlevelx, \atAlevely, 0.3*\atomtranw, 0.3*\atomtranh, 0, gray2, bcol)
\draw[dash pattern={on \padint off \padint on \padint off \padint on \padint}](\atAx, \atAy-0.3*\pad)--(\atAlevelx-0.3*\pad ,\atAlevely +0.3*\pad );
\draw[dash pattern={on \padint off \padint on \padint off \padint on \padint}](\atAx, \atAy+ 0.3*\pad)--(\atAlevelx-0.3*\pad ,\atAlevely + 0.3*\atomtranh + 0.3*\pad );
\draw[dashed,dash pattern=on \padint off \padint, gray2](\atAlevelx+ 0.5*0.3* \atomtranw, \atAlevely +  0.5*0.3*\atomtranh) ellipse ( 0.25*\atomtranw cm and 0.25*\atomtranh cm);
\draw[thick,bcol](\atAlevelx+ 0.5*0.3* \atomtranw, \atAlevely+ 0.3*\atomtranh ) circle (0.05);

\myatomTransition.two_levels(\atBlevelx, \atBlevely, 0.3*\atomtranw, \rbfac * 0.3*\atomtranh, 0, gray2, rcol)
\draw[dash pattern={on \padint off \padint on \padint off \padint on \padint}](\atBx, \atBy +0.0*\pad)--(\atBlevelx-0.3*\pad ,\atBlevely - 0.3*\pad );
\draw[dash pattern={on \padint off \padint on \padint off \padint on \padint}](\atBx, \atBy+ 0.8*\pad)--(\atBlevelx-0.3*\pad ,\atBlevely + 0.3*\rbfac*\atomtranh - 0.3*\pad );
\draw[dashed,dash pattern=on \padint off \padint, gray2](\atBlevelx+ 0.5*0.3* \atomtranw, \atBlevely +  \rbfac*0.5*0.3*\atomtranh) ellipse ( 0.25*\atomtranw cm and 0.25*\atomtranh *\rbfac*1.3 cm);
\draw[thick,rcol](\atBlevelx+ 0.5*0.3* \atomtranw, \atBlevely+\rbfac*0.3*\atomtranh ) circle (0.05);

\myatomTransition.two_levels(\atClevelx, \atClevely, 0.3*\atomtranw, \rbfac * 0.3*\atomtranh, 0, gray2, rcol)
\draw[dash pattern={on \padint off \padint on \padint off \padint on \padint}](\atCx, \atCy-0.3*\pad)--(\atClevelx-0.3*\pad ,\atClevely - 0.3*\pad );
\draw[dash pattern={on \padint off \padint on \padint off \padint on \padint}](\atCx, \atCy+ 0.3*\pad)--(\atClevelx-0.3*\pad ,\atClevely + 0.3*\rbfac*\atomtranh - 0.3*\pad );
\draw[dashed,dash pattern=on \padint off \padint, gray2](\atClevelx+ 0.5*0.3* \atomtranw, \atClevely +  \rbfac*0.5*0.3*\atomtranh) ellipse ( 0.25*\atomtranw cm and 0.25*\atomtranh *\rbfac*1.3 cm);
\draw[thick,rcol](\atClevelx+ 0.5*0.3* \atomtranw , \atClevely + 0.3*\rbfac*\atomtranh ) circle (0.05);

\end{scope}
\end{tikzpicture}
\caption{ \justifying Schematic plot of a simultaneous three-photon-three-atom process where the atoms
are jointly excited. The
 $A$-species atom (represented by the large ball) absorbs a laser $\mathfrak{L}_1$ photon (blue line), 
 while each of  two $B$-species atoms absorbs  a laser $\mathfrak{L}_2$ photon (red  line). }
 \label{fig_3p3a2L}
\end{figure}

The transition rate for an $m$-atom system, denoted as $W_{mpma}$, 
is typically low under moderate laser intensity due to
its nature as a high-order QED  process.
 However, within this
 atomic gas, a specific $A$-species atom, designated as the $A_o$ atom,
 can form a vast number of $m$-atom systems with 
numerous $B$-species atoms. According to  quantum mechanical principles,
all these $m$-atom systems engage in the MPMA process 
in parallel,  resulting in a significant enhancement of the total transition rate, $W_{a_o}$, for the $A_o$ atom.
 Let $N_{b\mathfrak{o}}$ represent the total number 
of $B$-species atoms  capable of forming an $m$-atom system with the $A_o$ atom.
 The total number of $m$-atom systems involving the $A_o$ atom, denoted 
by $\mathfrak{N}_{a_o}$, is approximated (roughly) as the combinatorial number \cite{yu2}:
 $C^{N_{b\mathfrak{o}}}_{m-1} = \frac{N_{b\mathfrak{o}}(N_{b\mathfrak{o}}-1)\dots (N_{b\mathfrak{o}}-m+1)}{(m-1)(m-2) \dots 1}
 \approx N_{b\mathfrak{o}}^{m-1}/(m-1)! $. Thus, the total transition rate can be expressed as:
 \begin{equation}
W_{a_o} = \mathfrak{N}_{a_o}  W_{mpma} \approx  \frac{N_{b\mathfrak{o}}^{m-1}}{(m-1)!} W_{mpma}.
\label{eq:W_ao}
\end{equation}
For instance, with $N_{b\mathfrak{o}} \approx 10^{12}$ and $m=8$, $\mathfrak{N}_{a_o}$
 can reach values as high as $10^{91}$ in principle, signifying 
 an exceptional enhancement factor. This yields a considerable total 
 transition rate for the $A_o$ atom.
  
This enhancement can be understood intuitively: in science and technology, it 
is frequently observed that while the signal from a single sample may be faint, 
a sufficiently large number of samples can generate a significant total signal. 
In typical atomic transition events, the number of samples corresponds to 
the number of atoms in the system, with the practical limit generally 
being below Avogadro's number, which limits the maximum possible 
enhancement. However, in this multiatom process, the number of `samples' is 
determined by a combinatorial factor rather than merely the number of 
atoms, enabling the total number to become exceptionally large.

In a homogeneous atomic gas, the value of  $N_{b\mathfrak{o}}$
is determined by $N_{b\mathfrak{o}} \approx \rho_b\, l_{mpma}^3 $,
 where $\rho_b$ represents the atomic density of  $B$-species atoms and $l_{mpma}$ 
 is a fundamental length which defines the maximum linear size of the $m$-atom 
 system enabling the simultaneous MPMA transition \cite{yu1,yu2}.  The length
 $l_{mpma}$ 
 is primarily governed by the uncertainty principle 
 in quantum mechanics and can be approximated as $l_{mpma}= \alpha c /2 (\Omega_{\mathfrak{L}_1} - \omega_a)$,
 where $c$ is the speed of light and $\alpha$ is a constant of order unity or less.

Further analysis reveals that this enhancement mechanism 
incorporates a regulatory near-saturation effect for $W_{a_o}$ \cite{yu2}. 
This implies that, while $W_{a_o}$ is a rapidly increasing function 
of both $N_{b\mathfrak{o}}$ and $m$ in its bare form, its maximum 
possible value remains below the scale of approximately $10^9$ s$^{-1}$. This 
regulation is facilitated by the self-tuning of $l_{mpma}$, which adjusts to 
reduce the values of $N_{b\mathfrak{o}}$ and, consequently, $W_{a_o}$, 
thereby ensuring compliance with relativistic causality \cite{yu2}.

 It is instructive
to estimate $W_{mpma}$, which can be calculated using
 perturbation theory and  approximated as follows \cite{yu2}:
\begin{equation}
 W_{mpma} \approx C {\Gamma_a    n_{\mathfrak{L_1}} \Omega_{\mathfrak{L_1}} } \frac{\Gamma^2  n^{m-1}_{\mathfrak{L_2}}\gamma_b^{m-1} \Omega^{m-1}_\mathfrak{L_2}} {(\Omega_\mathfrak{L_2} - \omega_b)^{2m}} [f(m)]^{2m} 
 \rho(E_f)|_{E_f= \varepsilon_e^a + (m-1)\varepsilon_e^b }.
 \label{eq:w2} 
\end{equation}
In this expression, $ C =  {(\Omega_\mathfrak{L_1}/\omega_a)}{(\Omega_\mathfrak{L_2}/\omega_b)^{m-1}}/ ({{2}^{4m-1}\pi^{3m-1} \hbar^2)} \approx {1}/({{2}^{4m-1}\pi^{3m-1}}\hbar^2)$;
 $\Gamma_a = {4\alpha_e \hbar}{ \omega^3_a |\langle e_a| \mathbf{{d}} | g_a \rangle|^2 }/{3 e^2 c^2}$ denotes
 an energy width parameter associated with the $E1$ transition of the $A$-species atom (with $\alpha_e$ being 
 the fine-structure constant, $e$ the elementary charge,  and $ \mathbf{{d}}$  the electric dipole moment operator);  
 $\gamma_b = {4\alpha_e} \omega^3_b |\langle e_b| \mathbf{{d}} | g_b \rangle|^2/ {3 e^2 c^2}$;
 and $\Gamma \approx \hbar\gamma_b$ represents an approximately averaged energy width parameter \cite{yu1, yu2}. 
 The term  $n_{\mathfrak{L_i}}(i=1,2)$ 
 represents the number of laser photons in laser $\mathfrak{L_i}$ within a  volume of $\lambda_i^3 = (2\pi c)^3 / 
 \Omega^3_\mathfrak{L_i}$. The function $f(m)$ takes values approximately in the range $(1/m,1)$, and $\rho(E_f)$ 
 denotes the density of states of the $m$-atom system at energy $E_f=\varepsilon_e^a + (m-1)\varepsilon_e^b $, 
 with $\varepsilon_e^a$ and 
 $\varepsilon_e^b$ representing the eigenenergies of states $| e_a \rangle$ and $| e_b \rangle$, respectively \cite{yu1,yu2}.
 
 By combining Equations~(\ref{eq:W_ao}) and (\ref{eq:w2}) and performing some algebraic manipulation, one can express
  $W_{a_o}$ in the following approximate form: 
 \begin{equation}
 W_{a_o}  \approx \frac{1}{8 \pi^2 \hbar^2} n_{\mathfrak{L_1}}  {\Gamma_a  \Omega_{\mathfrak{L}_1}}  \frac{\Gamma^2}{ (\Omega_{\mathfrak{L}_2}-\omega_b )^2}   \,\mathfrak{E}^{m-1}_{mpma}  \,
 \rho(E_f) .
 \label{eq:w3} 
\end{equation}
Here $ \mathfrak{E}^{m-1}_{mpma}$ denotes a composite enhancement factor, given by:
\begin{equation}
\mathfrak{E}^{m-1}_{mpma} = \left [\frac{ n_{\mathfrak{L_2}} \gamma_b \: \Omega_{\mathfrak{L}_2}}{ 16 \pi^3 (\Omega_{\mathfrak{L}_2}-\omega_b )^2} \frac{[f(m)]^{\frac{2m}{m-1}}}
{[(m-1)!]^\frac{1}{m-1}}  N_{b\mathfrak{o}} \right ]^{m-1} .
 \end{equation}
For  typical atomic species, $\gamma_b$ is on the order
of $2\pi \times 10$  MHz,  while $\omega_b$ is on the order of $2\pi \times 5 \times 10^{14}$ Hz.
The detuning parameter $|\Omega_{\mathfrak{L}_2}-\omega_b|$ can be set
to around $2\pi \times 10$ GHz practically. 
The enhancement factor $\mathfrak{E}_{mpma}^{m-1}$ can be on the order of, or greater than, 
 $ ( 0.1\, n_{\mathfrak{L_2}} N_{b\mathfrak{o}}/ m )^{m-1}$.
 Even with a very small value of $n_{\mathfrak{L}_2}$, such as $10^{-4}$, which corresponds 
 to a weak laser intensity, a sufficiently large $N_{b\mathfrak{o}}$, on the order of
 $10^{12}$ or greater, 
 enables the enhancement factor $\mathfrak{E}_{mpma}^{m-1}$ to 
 reach an exceptionally large value when $m$ is substantial.

This MPMA process can be incorporated into an ultraweak atomic transition, such as 
the atomic absorption of an $E5$-photon,  to
enhance its transition rate \cite{yu2}. Consider a similar $m$-atom system,
where the atomic transitions of the $B$-species atoms remain the same, but the 
transition of the $A$-species atom is replaced by the ultraweak  $E5$-photon atomic 
transition. In this case,  
the role of the laser $\mathfrak{L}_1$ is substituted by a flux of
$E5$ photons. 
The frequencies of the involved photons must satisfy the condition of 
overall energy conservation for the joint process, which is expressed as:
\begin{equation}
\hbar \Omega_{\scalebox{0.5}{$E$}_5} + (m-1) \hbar \Omega_{\mathfrak{L}_2} = \hbar \omega_{a, {\scalebox{0.5}{$E$}_5}} + (m-1) \hbar \omega_b,
\end{equation}
where $ \Omega_{\scalebox{0.5}{$E$}_5} $ denotes the frequency of the $E5$ photons and $\omega_{a, {\scalebox{0.5}{$E$}_5}}$ denotes the 
transition frequency for the $E5$ transition of the $A$-species atom. The transition rate for
a specific $A_o$ atom in an atomic gas can be approximated as analogous to Eq.~(\ref{eq:w2}): 
\begin{equation}
W_{a_o, {\scalebox{0.5}{$E$}_5}}  \approx \frac{1}{8 \pi^2 \hbar^2} { n_{\scalebox{0.5}{$E$}_5}   \Gamma_{a,  n_{\scalebox{0.5}{$E$}_5} }  \Omega_{\scalebox{0.5}{$E$}_5} } \frac{\Gamma^2}{ (\Omega_\mathfrak{L_2}-\omega_b )^2}  \, \mathfrak{E}^{m-1}_{mpma}\,\rho(E_f) .
\label{eq:wAo}
\end{equation}
In this expression, $n_{\scalebox{0.5}{$E$}_5}$ denotes the number of 
$E5$ photons in a volume of $\lambda_{\scalebox{0.5}{$E$}_5}^3 = (2 \pi c)^3 / 
 \Omega^3_{\scalebox{0.5}{$E$}_5}$. $ \Gamma_{a, {\scalebox{0.5}{$E$}_5}}$ is an energy
  width parameter 
associated with
the ultraweak $E5$ transition of the $A$-species atom, 
$\Gamma_{a, {\scalebox{0.5}{$E$}_5}} \sim \alpha_e \hbar \omega_{a, {\scalebox{0.5}{$E$}_5}} |\langle e_{a,{\scalebox{0.5}{$E$}_5}}| {r^5 Y_5} | g_a \rangle|^2/\lambda^{10}_{\scalebox{0.5}{$E$}_5}  $ 
($r$ is the radius of the electron, $Y$ is the spherical harmonic 
function, and $ |e_{a,{\scalebox{0.5}{$E$}_5}}\rangle$ is the corresponding
excited state), which could be  $10^{-32}$ smaller than the  width parameter $\Gamma_a $ for 
the $E1$ transition.
However, 
the enhancement factor $\mathfrak{E}^{m-1}_{mpma}$ can be made to
sufficiently large to overcome the smallness of $\Gamma_{a, {\scalebox{0.5}{$E$}_5}}$, allowing 
 the transition rate $W_{a_o, {\scalebox{0.5}{$E$}_5}}$  to reach a detectable level.

\begin{figure}
\captionsetup{width=.90\linewidth}
\begin{tikzpicture}
\begin{scope}[xshift=0, scale=1.0]

\mygravitonlineRatio.apply(\wx -3.5*\linexinc, -3*\wint -3.5*\lineyinc, \longline*\wsz, -2*\wint+\longline*\wsz, \asz, webblue, \large)
\mygravitonlineRatio.apply(\wx -2.5*\linexinc, -3*\wint -2.5*\lineyinc, \longline*\wsz, -2*\wint+\longline*\wsz, \asz, webblue, \large)
\mygravitonlineRatio.apply(\wx -1.5*\linexinc, -3*\wint -1.5*\lineyinc, \longline*\wsz, -2*\wint+\longline*\wsz, \asz, webblue, \large)
\myphotonlineRatio.apply(\wx, -3*\wint, \longline*\wsz, -2*\wint+ \longline*\wsz, \asz, redgold, 1.4*\large)
\myphotonlineRatio.apply(\wx + \linexinc, -3*\wint + \lineyinc, \longline*\wsz, -2*\wint+\longline*\wsz, \asz, redgold, 1.4*\large)
\myphotonlineRatio.apply(\wx + 2*\linexinc, -3*\wint + 2*\lineyinc, \longline*\wsz, -2*\wint+\longline*\wsz, \asz, redgold, 1.4*\large)
\myphotonlineRatio.apply(\wx + 3*\linexinc, -3*\wint + 3*\lineyinc, \longline*\wsz, -2*\wint+\longline*\wsz, \asz, redgold, 1.4*\large)
\myphotonlineRatio.apply(\wx + 4*\linexinc, -3*\wint + 4*\lineyinc, \longline*\wsz, -2*\wint+\longline*\wsz, \asz, redgold, 1.4*\large)
\myphotonlineRatio.apply(\wx + 5*\linexinc, -3*\wint + 5*\lineyinc, \longline*\wsz, -2*\wint+\longline*\wsz, \asz, redgold, 1.4*\large)

\node[circle, shading=ball, ball color=unexcited-rcol,  minimum size=2pt, scale=1.25*\atomsize,
text=white] (ball) at(\atAx,  \atAy){} ;
\node[circle, shading=ball, ball color=unexcited-rcol,  minimum size=2pt, scale=1.*\atomsize,
text=white] (ball) at(\atBx,  \atBy){} ;
\node[circle, shading=ball, ball color=unexcited-rcol,  minimum size=2pt, scale=1.*\atomsize,
text=white] (ball) at(\atCx,  \atCy){} ;
\node[circle, shading=ball, ball color=unexcited-rcol,  minimum size=2pt, scale=1.*\atomsize,
text=white] (ball) at(\atDx,  \atDy){} ;

\myatomTransition.two_levels(\atAlevelx, \atAlevely, 0.3*\atomtranw, 0.3*\atomtranh, 0, gray2, bcol)
\draw[dash pattern={on \padint off \padint on \padint off \padint on \padint}](\atAx, \atAy-0.3*\pad)--(\atAlevelx-0.3*\pad ,\atAlevely +0.3*\pad );
\draw[dash pattern={on \padint off \padint on \padint off \padint on \padint}](\atAx, \atAy+ 0.3*\pad)--(\atAlevelx-0.3*\pad ,\atAlevely + 0.3*\atomtranh + 0.3*\pad );
\draw[dashed,dash pattern=on \padint off \padint, gray2, -stealth](\atAlevelx+ 0.5*0.3*\atomtranw, \atAlevely )--(\atAlevelx+ 0.5*0.3*\atomtranw, \atAlevely +  0.3*\atomtranh);

\draw[dashed,dash pattern=on \padint off \padint, gray2](\atAlevelx+ 0.5*0.3* \atomtranw, \atAlevely +  0.5*0.3*\atomtranh) ellipse ( 0.25*\atomtranw cm and 0.25*\atomtranh cm);
\draw[thick,gray2](\atAlevelx+ 0.5*0.3 *\atomtranw, \atAlevely ) circle (0.05);

\myatomTransition.two_levels(\atDlevelx, \atDlevely, 0.3*\atomtranw, \rbfac * 0.3*\atomtranh, 0, gray2, rcol)
\draw[dash pattern={on \padint off \padint on \padint off \padint on \padint}](\atDx, \atDy-0.0*\pad)--(\atDlevelx-0.3*\pad ,\atDlevely - 0.3*\pad );
\draw[dash pattern={on \padint off \padint on \padint off \padint on \padint}](\atDx, \atDy+ 0.8*\pad)--(\atDlevelx-0.3*\pad ,\atDlevely + 0.3*\rbfac*\atomtranh - 0.3*\pad );
\draw[dashed,dash pattern=on \padint off \padint, gray2, -stealth](\atDlevelx+ 0.5*0.3* \atomtranw, \atDlevely )--(\atDlevelx+ 0.5*0.3*\atomtranw, \atDlevely +  0.3*\rbfac*\atomtranh);
\draw[dashed,dash pattern=on \padint off \padint, gray2](\atDlevelx+ 0.5*0.3* \atomtranw, \atDlevely +  \rbfac*0.5*0.3*\atomtranh) ellipse ( 0.25*\atomtranw cm and 0.25*\atomtranh *\rbfac*1.3 cm);
\draw[thick,gray2](\atDlevelx+ 0.5*0.3* \atomtranw, \atDlevely ) circle (0.05);

\end{scope}
\begin{scope}[xshift=100, scale=1.0]
\tikzmath{ \lineyinc= \lineyinc * 1.2;}
\mygravitonlineRatio.apply(\wx -1.2*\linexinc+\lineshift, -3*\wint -1.2*\lineyinc, \longline*\wsz, -2*\wint+\longline*\wsz, \asz, webblue, \large)
\mygravitonlineRatio.apply(\wx -2.2*\linexinc+\lineshift, -3*\wint -2.2*\lineyinc, \longline*\wsz, -2*\wint+\longline*\wsz, \asz, webblue, \large)

\myphotonlineRatio.apply(\wx +\lineshift, -3*\wint, \longline*\wsz, -2*\wint+ \longline*\wsz, \asz, redgold, 1.4*\large)
\myphotonlineRatio.apply(\wx + 1.0*\linexinc +\lineshift, -3*\wint + 1.0*\lineyinc, \longline*\wsz, -2*\wint+\longline*\wsz, \asz, redgold, 1.4*\large)
\myphotonlineRatio.apply(\wx + 2.0*\linexinc +\lineshift, -3*\wint + 2.0*\lineyinc, \longline*\wsz, -2*\wint+\longline*\wsz, \asz, redgold, 1.4*\large)

\node[circle, shading=ball, ball color=bcol,  minimum size=2pt, scale=1.25*\atomsize,
text=white] (ball) at(\atAx,  \atAy){} ;
\node[circle, shading=ball, ball color=rcol,  minimum size=2pt, scale=1.*\atomsize,
text=white] (ball) at(\atBx,  \atBy){} ;
\node[circle, shading=ball, ball color=rcol,  minimum size=2pt, scale=1.*\atomsize,
text=white] (ball) at(\atCx,  \atCy){} ;
\node[circle, shading=ball, ball color= rcol, minimum size=2pt, scale=1.*\atomsize,
text=white] (ball) at(\atDx,  \atDy){} ;
\myatomTransition.two_levels(\atAlevelx, \atAlevely, 0.3*\atomtranw, 0.3*\atomtranh, 0, gray2, bcol)
\draw[dash pattern={on \padint off \padint on \padint off \padint on \padint}](\atAx, \atAy-0.3*\pad)--(\atAlevelx- 0.3*\pad ,\atAlevely +-.3*\pad );
\draw[dash pattern={on \padint off \padint on \padint off \padint on \padint}](\atAx, \atAy+ 0.3*\pad)--(\atAlevelx-0.3*\pad ,\atAlevely + 0.3*\atomtranh + 0.3*\pad );
\draw[dashed,dash pattern=on \padint off \padint, gray2](\atAlevelx+ 0.5*0.3* \atomtranw, \atAlevely +  0.5*0.3*\atomtranh) ellipse ( 0.25*\atomtranw cm and 0.25*\atomtranh cm);
\draw[thick,blue](\atAlevelx+ 0.5*0.3* \atomtranw, \atAlevely+ 0.3*\atomtranh ) circle (0.05);

\myatomTransition.two_levels(\atDlevelx, \atDlevely, 0.3*\atomtranw, \rbfac * 0.3*\atomtranh, 0, gray2, rcol)
\draw[dash pattern={on \padint off \padint on \padint off \padint on \padint}](\atDx, \atDy +0.0*\pad)--(\atDlevelx-0.3*\pad ,\atDlevely - 0.3*\pad );
\draw[dash pattern={on \padint off \padint on \padint off \padint on \padint}](\atDx, \atDy+ 0.8*\pad)--(\atDlevelx-0.3*\pad ,\atDlevely + 0.3*\rbfac*\atomtranh - 0.3*\pad );
\draw[dashed,dash pattern=on \padint off \padint, gray2](\atDlevelx+ 0.5*0.3* \atomtranw, \atDlevely +  \rbfac*0.5*0.3*\atomtranh) ellipse ( 0.25*\atomtranw cm and 0.25*\atomtranh *\rbfac*1.3 cm);
\draw[thick,rcol](\atDlevelx+ 0.5*0.3* \atomtranw, \atDlevely+\rbfac*0.3*\atomtranh ) circle (0.05);

\end{scope}
\end{tikzpicture}
\caption{ \justifying Schematic  illustration of a graviton-absorption atomic process involving simultaneous joint excitations of a four-atom system. The
 $A$-species atom (represented by the large ball) absorbs a graviton (curly line), 
 while each of  three $B$-species atoms absorbs a laser photon (wavy line). 
  }
 \label{fig_4p1g4a}
\end{figure}
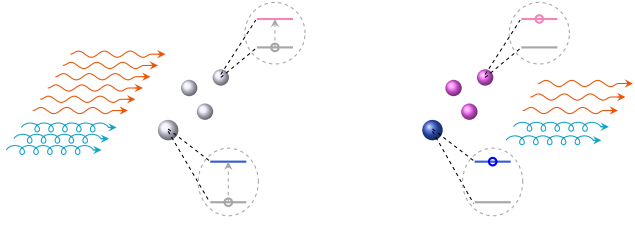

We can extend this MPMA enhancement to atomic transitions involving graviton absorption, 
in a manner analogous to the ultraweak atomic 
transition involving an $E5$ photon.
 Consider a flux of gravitons with
frequency $\Omega_{gr}$. A suitable $A$-species atom and $m-1$ $B$-species 
atoms can be selected, along with a laser of frequency $\Omega_{\mathfrak{L}_2}$,
 such that the following energy conservation condition is satisfied:

\begin{equation}
\hbar \Omega_{gr}  + (m-1) \hbar \Omega_{\mathfrak{L}_2}=  \hbar \omega_{a,gr} + (m-1) \hbar \omega_b,
\label{eq:gr_energy}
\end{equation}
where $\omega_{a,gr}$ denotes the 
transition frequency for the graviton-absorptive transition of the $A$-species atom.
In this scenario, an analogous MPMA process of this 
$m$-atom system can occur.  
The $A$-species atom absorbs the graviton while, simultaneously, 
the $m-1$ $B$-species atoms undergo transitions by  each absorbing a laser photon
(see Fig. \ref{fig_4p1g4a} for an example with $m=4$).

       \tikzmath{ \atAy= -16*\wint; \atBx= \atAx + 0.4*\atgap;
      \atBy= \atAy + 0.8*\atgap; \atCx= \atAx + 0.7*\atgap; \atCy= \atAy + 0.35*\atgap;
      \atDx= \atAx + 1.0*\atgap; \atDy= \atAy + 1.0*\atgap;  \atDlevelx= \atDx +0.6*\levelgap;
       \atDlevely= \atDy + 0.5*\levelgap;  \atAlevelx= \atAx +0.7*\levelgap;
       \atAlevely= \atAy - 1.2*\levelgap;}

Now, consider an atomic gas undergoing graviton absorption. 
The overall transition rate of a specific $A$-species atom in the gas
can be  analyzed in a manner similar to previous treatments and can 
be approximated in a form analogous to 
 Eq.~(\ref{eq:w2}): 
\begin{equation}
W_{a_o,gr} \approx \frac{1}{8 \pi^2 \hbar^2} n_{gr}   \Gamma_{a,gr}  \Omega_{gr } \frac{\Gamma^2}{ (\Omega_\mathfrak{L_2}-\omega_b )^2} \,  \mathfrak{E}^{m-1}_{mpma}\, \rho(E_f).
\label{eq:wAogr}
\end{equation}
Here, $n_{gr}$ denotes the number of 
gravitons in a volume of $\lambda_{gr}^3 = (2\pi c)^3 / 
 \Omega^3_{gr}$, and $\Gamma_{a,gr} $ represents an energy width parameter 
associated with the graviton-absorption transition of the $A$-species atom.
Formally, the interaction operator (density) 
between the atom and a weak gravitational
field can be approximated  as \cite{BoughnHydrogenGraviton2006, Pitelli2021angular}:
\begin{equation}
H_{ge} \approx \frac {m_e}{2} R_{0i0j}(t,\mathbf{x})x^i x^j .
\end{equation}
In this expression,  $m_e$ is the electron mass and $x^i$ denotes the Fermi
normal coordinate in the atom's rest space, with $i,j$ as spatial indices. 
The term $R$ refers to 
the curvature tensor operator of the gravitational
field.   The energy width parameter $\Gamma_{a,gr}$ $\sim $
$ {G m^2_e  \omega_{a,gr}}|\langle e_{a,gr}| x^i x^j | g_a \rangle|^2/c \lambda^4_{gr} $, where $G$ is the gravitational constant and $| e_{a,gr}\rangle$
is the corresponding excited state,  is roughly
$10^{-50}$ smaller than width 
parameter $ \Gamma_a$ for the $E1$ transition.
 However, owing to the potentially enormous value of  $\mathfrak{E}^{m-1}_{mpma}$, 
the transition rate $W_{a_o,gr}$ can reach detectable level despite
the extremely small magnitude of  ${\Gamma_{a,gr}}$.

In the graviton-absorption process, the incoming flux of natural 
gravitons may sometimes be extremely low—potentially below one 
graviton per minute per square centimeter. In such cases, 
the primary interest could be the absorption probability of a graviton.
If the frequency of an incoming graviton satisfies the energy 
condition (Eq.~(\ref{eq:gr_energy})),
enabling the joint MPMA process to occur,  the 
absorption probability is significantly enhanced. 
Although the probability of a single $m$-atom system 
absorbing an incoming graviton remains exceedingly small,
the existence of a vast number of such systems, 
 all capable of absorbing gravitons in parallel,
 allows the total absorption probability for 
  the entire sample to approach unity.
 
A graviton carries an intrinsic angular momentum of $2 \hbar$,
 which generally requires that for an $A$-species atom to absorb 
 a graviton, the corresponding atomic transition must involve 
 a change of two units of angular momentum—analogous to an $E2$
photoabsorption transition. Once excited, the $A$-species atom 
can often emit an $E1$ photon by transitioning from the excited 
state to a distinct lower-energy excited state, 
rather than the ground state.
 This 
$E1$ photon, distinguished by its characteristic frequency, 
acts as the observable signature of graviton absorption.

An alternative class of material systems exists in which graviton 
absorption occurs via quantum transitions between states that 
are not eigenstates of the angular momentum operator. Consequently, 
the conventional selection rules tied to angular momentum quantum numbers do not apply. 
These systems consist of ion-doped crystals,
 where the doped ions \textemdash such as rare-earth elements 
  like Eu$^{3+}$, Nd$^{3+}$, and Pr$^{3+}$\textemdash 
serve as  active sites for graviton-involved multiphoton-multiparticle
transitions,  also referred to as the (generalized) MPMA process. 
Compared to atomic gases, ion-doped crystals
can be more beneficial for graviton absorption. In these solid-state systems,
the doped ions can easily achieve high densities, reaching 
$ 10^{18}$ ions/cm$^{-3}$ even at very low doping rates.
To implement MPMA-assisted graviton absorption in the crystal, two quantum 
transitions within the same ion species can be selected, eliminating 
the need for two distinct species.
 For instance, in a single ion species, the 
transition between the ground state and the first excited state  can 
be used for laser photon absorption, mirroring the transitions of 
$B$-species atoms in atomic gases. Meanwhile, another transition—between 
the ground state and a second excited state, well-separated in energy from the first 
excited state—can facilitate graviton absorption, analogous to the transitions in 
$A$-species atoms in atomic gases. 
 
An important factor concerning  MPMA process in ion-doped crystals is
the inhomogeneous 
broadening of the ions' optical transitions, denoted by $\Gamma_{inh}$.  
 In some systems,
$\Gamma_{inh}/\hbar $ can be as large as  $ 10^2$ GHz or more; however, it 
 can be reduced to a few 
hundred MHz in certain systems \cite{macfarlane1992inhomogeneous, 
macfarlane1998optical, chukalina2000fine, liu2006spectroscopic} with 
low-doping rates. Additionally, the natural linewidths of rare-earth ions’ 
optical transitions, 
 $\Gamma_{ion}/\hbar$, can be very small,  on the order 
of several kHz or less in some cases. 
 Due to  inhomogeneous broadening, 
not all possible  $m$-ion systems within
a length scale $l_{mpma}$ contribute equally, as their excitation 
energies exhibit a spread of approximately
$m^\eta  \Gamma_{inh}$, where the exponent $\eta$ ranges between 0.5 and 1. 
As a result, only a fraction of these $m$-ion systems
—roughly ${\Gamma_{ion}}/{m^\eta  \Gamma_{inh}}$ or larger—effectively 
participates in the MPMA process, 
introducing a reduction factor into the transition rate. 
 However, this reduction is 
  readily offset by the vastly larger combinatorial number 
  of possible $m$-ion systems, and the
 graviton absorption process can still reach a near-saturation regime.
 In this regime, $l_{mpma}$
self-adjusts to constrain what would otherwise be an unphysically large transition rate.
Another relevant factor is the thermal broadening of the excited
states of the ions, and the crystal can be cooled 
to cryogenic temperatures of a few kelvins to reduce 
the thermal broadening width. 
 
For graviton absorption to occur, a corresponding graviton source
is required. The possibility of enhanced graviton emission in atomic gases 
or ion-doped crystals is explored in a separate study. If such enhancement
is realized, gravitons could be generated with well-controlled frequencies, 
offering a more tunable approach to graviton production.
   
Natural high-frequency graviton sources 
include solar gravitons \cite{Weinberg1965, Hu2021high, 
Garcia-Cely2024} and relic gravitons. 
The estimated power of 
solar graviton emission is approximately $10^5$ W 
in the optical frequency range \cite{Garcia-Cely2024}, corresponding to a flux of 
about 10 gravitons per square centimeter per day at Earth's 
solar distance. Given the broad frequency distribution of solar gravitons,  enhancing
the detection bandwidth for graviton absorption becomes 
an essential consideration. In ion-doped 
crystals, the detection bandwidth depends on the spectral 
broadening of the excited levels involved in the MPMA process.
Denoting the total broadening--including both inhomogeneous and 
thermal broadening--of an ion's excited level as $\Gamma_{tot}$,
the total excitation energy of an $m$-ion system exhibits a spectral spread 
 of approximately $m^\gamma \Gamma_{tot}$,  which defines the detection bandwidth.
A larger $\Gamma_{tot}$ thus increases the detection bandwidth, though it may 
simultaneously reduce the number of $m$-ion systems capable of absorbing 
gravitons, thereby influencing the MPMA process. Nevertheless, as 
previously argued, this reduction can be counterbalanced, and
the MPMA process  can still reach a near-saturation regime. Additionally, 
under the same laser field, an ion-doped
 crystal can respond to solar gravitons across different 
frequency regimes by 
adapting different $m$ values. Consequently, 
considering multiple $m$ values,  the total detection bandwidth 
could reach $10^3$ GHz or beyond, making the detection of solar
 gravitons more feasible.

It is interesting to explore the possibility of extending 
graviton absorption into 
the infrared and ultraviolet frequency ranges. Expanding detection into the 
ultraviolet range appears relatively 
straightforward and can be achieved using methods analogous to those 
employed in the optical frequency range. However, extending detection 
  into the far-infrared range may present challenges and  requires further investigation.
  Broadening the detectable frequency range in this manner could 
  significantly expand the observational window for studying 
  astronomical gravitons, opening new opportunities for exploration and discovery.

In summary, we propose that atomic graviton absorption can 
be readily observed with the assistance of an MPMA process. 
Detecting this quantum 
phenomenon is not only essential for understanding the nature 
of gravitational waves but also fundamental to advancing 
our knowledge of the quantum framework itself.


\bibliographystyle{apsrev4-1}

\bibliography{gravitonbib}

\begin{thebibliography}{36}%
\makeatletter
\providecommand \@ifxundefined [1]{%
 \@ifx{#1\undefined}
}%
\providecommand \@ifnum [1]{%
 \ifnum #1\expandafter \@firstoftwo
 \else \expandafter \@secondoftwo
 \fi
}%
\providecommand \@ifx [1]{%
 \ifx #1\expandafter \@firstoftwo
 \else \expandafter \@secondoftwo
 \fi
}%
\providecommand \natexlab [1]{#1}%
\providecommand \enquote  [1]{``#1''}%
\providecommand \bibnamefont  [1]{#1}%
\providecommand \bibfnamefont [1]{#1}%
\providecommand \citenamefont [1]{#1}%
\providecommand \href@noop [0]{\@secondoftwo}%
\providecommand \href [0]{\begingroup \@sanitize@url \@href}%
\providecommand \@href[1]{\@@startlink{#1}\@@href}%
\providecommand \@@href[1]{\endgroup#1\@@endlink}%
\providecommand \@sanitize@url [0]{\catcode `\\12\catcode `\$12\catcode
  `\&12\catcode `\#12\catcode `\^12\catcode `\_12\catcode `\%12\relax}%
\providecommand \@@startlink[1]{}%
\providecommand \@@endlink[0]{}%
\providecommand \url  [0]{\begingroup\@sanitize@url \@url }%
\providecommand \@url [1]{\endgroup\@href {#1}{\urlprefix }}%
\providecommand \urlprefix  [0]{URL }%
\providecommand \Eprint [0]{\href }%
\providecommand \doibase [0]{http://dx.doi.org/}%
\providecommand \selectlanguage [0]{\@gobble}%
\providecommand \bibinfo  [0]{\@secondoftwo}%
\providecommand \bibfield  [0]{\@secondoftwo}%
\providecommand \translation [1]{[#1]}%
\providecommand \BibitemOpen [0]{}%
\providecommand \bibitemStop [0]{}%
\providecommand \bibitemNoStop [0]{.\EOS\space}%
\providecommand \EOS [0]{\spacefactor3000\relax}%
\providecommand \BibitemShut  [1]{\csname bibitem#1\endcsname}%
\let\auto@bib@innerbib\@empty
\bibitem [{\citenamefont {Abbott}\ \emph
  {et~al.}(2016{\natexlab{a}})\citenamefont {Abbott} \emph
  {et~al.}}]{abbott2016observation}%
  \BibitemOpen
  \bibfield  {author} {\bibinfo {author} {\bibfnamefont {B.~P.}\ \bibnamefont
  {Abbott}} \emph {et~al.},\ }\href@noop {} {\bibfield  {journal} {\bibinfo
  {journal} {Physical review letters}\ }\textbf {\bibinfo {volume} {116}},\
  \bibinfo {pages} {061102} (\bibinfo {year} {2016}{\natexlab{a}})}\BibitemShut
  {NoStop}%
\bibitem [{\citenamefont {Abbott}\ \emph
  {et~al.}(2016{\natexlab{b}})\citenamefont {Abbott} \emph
  {et~al.}}]{abbott2016gw151226}%
  \BibitemOpen
  \bibfield  {author} {\bibinfo {author} {\bibfnamefont {B.~P.}\ \bibnamefont
  {Abbott}} \emph {et~al.},\ }\href@noop {} {\bibfield  {journal} {\bibinfo
  {journal} {Physical review letters}\ }\textbf {\bibinfo {volume} {116}},\
  \bibinfo {pages} {241103} (\bibinfo {year} {2016}{\natexlab{b}})}\BibitemShut
  {NoStop}%
\bibitem [{\citenamefont {Abbott}\ \emph {et~al.}(2017)\citenamefont {Abbott}
  \emph {et~al.}}]{abbott2017gw170817}%
  \BibitemOpen
  \bibfield  {author} {\bibinfo {author} {\bibfnamefont {B.~P.}\ \bibnamefont
  {Abbott}} \emph {et~al.},\ }\href@noop {} {\bibfield  {journal} {\bibinfo
  {journal} {Physical review letters}\ }\textbf {\bibinfo {volume} {119}},\
  \bibinfo {pages} {161101} (\bibinfo {year} {2017})}\BibitemShut {NoStop}%
\bibitem [{\citenamefont {Dyson}(2004)}]{dyson2004world}%
  \BibitemOpen
  \bibfield  {author} {\bibinfo {author} {\bibfnamefont {F.}~\bibnamefont
  {Dyson}},\ }\href@noop {} {\bibfield  {journal} {\bibinfo  {journal} {New
  York Review of Books}\ }\textbf {\bibinfo {volume} {51}} (\bibinfo {year}
  {2004})}\BibitemShut {NoStop}%
\bibitem [{\citenamefont {Dyson}(2013)}]{dyson2013graviton}%
  \BibitemOpen
  \bibfield  {author} {\bibinfo {author} {\bibfnamefont {F.}~\bibnamefont
  {Dyson}},\ }\href {\doibase 10.1142/S0217751X1330041X} {\bibfield  {journal}
  {\bibinfo  {journal} {International Journal of Modern Physics A}\ }\textbf
  {\bibinfo {volume} {28}},\ \bibinfo {pages} {1330041} (\bibinfo {year}
  {2013})}\BibitemShut {NoStop}%
\bibitem [{\citenamefont {Rothman}\ and\ \citenamefont
  {Boughn}(2006)}]{Rothman2006can}%
  \BibitemOpen
  \bibfield  {author} {\bibinfo {author} {\bibfnamefont {T.}~\bibnamefont
  {Rothman}}\ and\ \bibinfo {author} {\bibfnamefont {S.}~\bibnamefont
  {Boughn}},\ }\href {\doibase 10.1007/s10701-006-9081-9} {\bibfield  {journal}
  {\bibinfo  {journal} {Foundations of Physics}\ }\textbf {\bibinfo {volume}
  {36}},\ \bibinfo {pages} {1801} (\bibinfo {year} {2006})}\BibitemShut
  {NoStop}%
\bibitem [{\citenamefont {Boughn}\ and\ \citenamefont
  {Rothman}(2006)}]{BoughnHydrogenGraviton2006}%
  \BibitemOpen
  \bibfield  {author} {\bibinfo {author} {\bibfnamefont {S.}~\bibnamefont
  {Boughn}}\ and\ \bibinfo {author} {\bibfnamefont {T.}~\bibnamefont
  {Rothman}},\ }\href {\doibase 10.1088/0264-9381/23/20/006} {\bibfield
  {journal} {\bibinfo  {journal} {Classical and Quantum Gravity}\ }\textbf
  {\bibinfo {volume} {23}},\ \bibinfo {pages} {5839} (\bibinfo {year}
  {2006})}\BibitemShut {NoStop}%
\bibitem [{\citenamefont {Pitelli}\ and\ \citenamefont
  {Perche}(2021)}]{Pitelli2021angular}%
  \BibitemOpen
  \bibfield  {author} {\bibinfo {author} {\bibfnamefont {J.~P.~M.}\
  \bibnamefont {Pitelli}}\ and\ \bibinfo {author} {\bibfnamefont {T.~R.}\
  \bibnamefont {Perche}},\ }\href {\doibase 10.1103/PhysRevD.104.065016}
  {\bibfield  {journal} {\bibinfo  {journal} {Phys. Rev. D}\ }\textbf {\bibinfo
  {volume} {104}},\ \bibinfo {pages} {065016} (\bibinfo {year}
  {2021})}\BibitemShut {NoStop}%
\bibitem [{\citenamefont {Hu}\ and\ \citenamefont {Yu}(2021)}]{Hu2021high}%
  \BibitemOpen
  \bibfield  {author} {\bibinfo {author} {\bibfnamefont {J.}~\bibnamefont
  {Hu}}\ and\ \bibinfo {author} {\bibfnamefont {H.}~\bibnamefont {Yu}},\ }\href
  {\doibase 10.1140/epjc/s10052-021-09312-9} {\bibfield  {journal} {\bibinfo
  {journal} {The European Physical Journal C}\ }\textbf {\bibinfo {volume}
  {81}},\ \bibinfo {pages} {470} (\bibinfo {year} {2021})}\BibitemShut
  {NoStop}%
\bibitem [{\citenamefont {Carney}\ \emph {et~al.}(2024)\citenamefont {Carney},
  \citenamefont {Domcke},\ and\ \citenamefont
  {Rodd}}]{carney2024grvitonclassicalExploration}%
  \BibitemOpen
  \bibfield  {author} {\bibinfo {author} {\bibfnamefont {D.}~\bibnamefont
  {Carney}}, \bibinfo {author} {\bibfnamefont {V.}~\bibnamefont {Domcke}}, \
  and\ \bibinfo {author} {\bibfnamefont {N.~L.}\ \bibnamefont {Rodd}},\ }\href
  {\doibase 10.1103/PhysRevD.109.044009} {\bibfield  {journal} {\bibinfo
  {journal} {Physical Review D}\ }\textbf {\bibinfo {volume} {109}},\ \bibinfo
  {pages} {044009} (\bibinfo {year} {2024})}\BibitemShut {NoStop}%
\bibitem [{\citenamefont {Marletto}\ and\ \citenamefont
  {Vedral}(2017)}]{Marletto2017TwoMass}%
  \BibitemOpen
  \bibfield  {author} {\bibinfo {author} {\bibfnamefont {C.}~\bibnamefont
  {Marletto}}\ and\ \bibinfo {author} {\bibfnamefont {V.}~\bibnamefont
  {Vedral}},\ }\href {\doibase 10.1103/PhysRevLett.119.240402} {\bibfield
  {journal} {\bibinfo  {journal} {Physical Review Letters}\ }\textbf {\bibinfo
  {volume} {119}},\ \bibinfo {pages} {240402} (\bibinfo {year}
  {2017})}\BibitemShut {NoStop}%
\bibitem [{\citenamefont {Bose}\ \emph {et~al.}(2017)\citenamefont {Bose},
  \citenamefont {Mazumdar}, \citenamefont {Morley}, \citenamefont {Ulbricht},
  \citenamefont {Toro{\v s}}, \citenamefont {Paternostro}, \citenamefont
  {Geraci}, \citenamefont {Barker}, \citenamefont {Kim},\ and\ \citenamefont
  {Milburn}}]{SpinEntanglement}%
  \BibitemOpen
  \bibfield  {author} {\bibinfo {author} {\bibfnamefont {S.}~\bibnamefont
  {Bose}}, \bibinfo {author} {\bibfnamefont {A.}~\bibnamefont {Mazumdar}},
  \bibinfo {author} {\bibfnamefont {G.~W.}\ \bibnamefont {Morley}}, \bibinfo
  {author} {\bibfnamefont {H.}~\bibnamefont {Ulbricht}}, \bibinfo {author}
  {\bibfnamefont {M.}~\bibnamefont {Toro{\v s}}}, \bibinfo {author}
  {\bibfnamefont {M.}~\bibnamefont {Paternostro}}, \bibinfo {author}
  {\bibfnamefont {A.~A.}\ \bibnamefont {Geraci}}, \bibinfo {author}
  {\bibfnamefont {P.~F.}\ \bibnamefont {Barker}}, \bibinfo {author}
  {\bibfnamefont {M.}~\bibnamefont {Kim}}, \ and\ \bibinfo {author}
  {\bibfnamefont {G.}~\bibnamefont {Milburn}},\ }\href {\doibase
  10.1103/PhysRevLett.119.240401} {\bibfield  {journal} {\bibinfo  {journal}
  {Physical Review Letters}\ }\textbf {\bibinfo {volume} {119}},\ \bibinfo
  {pages} {240401} (\bibinfo {year} {2017})}\BibitemShut {NoStop}%
\bibitem [{\citenamefont {Parikh}\ \emph {et~al.}(2020)\citenamefont {Parikh},
  \citenamefont {Wilczek},\ and\ \citenamefont {Zahariade}}]{Wilczek1}%
  \BibitemOpen
  \bibfield  {author} {\bibinfo {author} {\bibfnamefont {M.}~\bibnamefont
  {Parikh}}, \bibinfo {author} {\bibfnamefont {F.}~\bibnamefont {Wilczek}}, \
  and\ \bibinfo {author} {\bibfnamefont {G.}~\bibnamefont {Zahariade}},\
  }\href@noop {} {\bibfield  {journal} {\bibinfo  {journal} {International
  Journal of Modern Physics D}\ }\textbf {\bibinfo {volume} {29}},\ \bibinfo
  {pages} {2042001} (\bibinfo {year} {2020})},\ \bibinfo {note} {award-Winning
  Essay}\BibitemShut {NoStop}%
\bibitem [{\citenamefont {Parikh}\ \emph {et~al.}(2021)\citenamefont {Parikh},
  \citenamefont {Wilczek},\ and\ \citenamefont {Zahariade}}]{Wilczek2}%
  \BibitemOpen
  \bibfield  {author} {\bibinfo {author} {\bibfnamefont {M.}~\bibnamefont
  {Parikh}}, \bibinfo {author} {\bibfnamefont {F.}~\bibnamefont {Wilczek}}, \
  and\ \bibinfo {author} {\bibfnamefont {G.}~\bibnamefont {Zahariade}},\ }\href
  {\doibase 10.1103/PhysRevLett.127.081602} {\bibfield  {journal} {\bibinfo
  {journal} {Physical Review Letters}\ }\textbf {\bibinfo {volume} {127}},\
  \bibinfo {pages} {081602} (\bibinfo {year} {2021})}\BibitemShut {NoStop}%
\bibitem [{\citenamefont {He}\ and\ \citenamefont
  {Zhang}(2022)}]{Zhangbaocheng}%
  \BibitemOpen
  \bibfield  {author} {\bibinfo {author} {\bibfnamefont {F.}~\bibnamefont
  {He}}\ and\ \bibinfo {author} {\bibfnamefont {B.}~\bibnamefont {Zhang}},\
  }\href {\doibase 10.1103/PhysRevD.105.106019} {\bibfield  {journal} {\bibinfo
   {journal} {Phys. Rev. D}\ }\textbf {\bibinfo {volume} {105}},\ \bibinfo
  {pages} {106019} (\bibinfo {year} {2022})}\BibitemShut {NoStop}%
\bibitem [{\citenamefont {Kanno}\ \emph {et~al.}(2021)\citenamefont {Kanno},
  \citenamefont {Soda},\ and\ \citenamefont {Tokuda}}]{Kanno2021}%
  \BibitemOpen
  \bibfield  {author} {\bibinfo {author} {\bibfnamefont {S.}~\bibnamefont
  {Kanno}}, \bibinfo {author} {\bibfnamefont {J.}~\bibnamefont {Soda}}, \ and\
  \bibinfo {author} {\bibfnamefont {J.}~\bibnamefont {Tokuda}},\ }\href
  {\doibase 10.1103/PhysRevD.103.044017} {\bibfield  {journal} {\bibinfo
  {journal} {Physical Review D}\ }\textbf {\bibinfo {volume} {103}},\ \bibinfo
  {pages} {044017} (\bibinfo {year} {2021})}\BibitemShut {NoStop}%
\bibitem [{\citenamefont {Tobar}\ \emph {et~al.}(2024)\citenamefont {Tobar},
  \citenamefont {Manikandan}, \citenamefont {Beitel},\ and\ \citenamefont
  {Pikovski}}]{tobar2024detecting}%
  \BibitemOpen
  \bibfield  {author} {\bibinfo {author} {\bibfnamefont {G.}~\bibnamefont
  {Tobar}}, \bibinfo {author} {\bibfnamefont {S.~K.}\ \bibnamefont
  {Manikandan}}, \bibinfo {author} {\bibfnamefont {T.}~\bibnamefont {Beitel}},
  \ and\ \bibinfo {author} {\bibfnamefont {I.}~\bibnamefont {Pikovski}},\
  }\href {\doibase 10.1038/s41467-024-07229-1} {\bibfield  {journal} {\bibinfo
  {journal} {Nature Communications}\ }\textbf {\bibinfo {volume} {15}},\
  \bibinfo {pages} {7229} (\bibinfo {year} {2024})}\BibitemShut {NoStop}%
\bibitem [{Gco()}]{Gcoupling}%
  \BibitemOpen
  \href@noop {} {}\bibinfo {note} {Here we refer to the ratio of the
  gravitational force to the electromagnetic force between two
  electrons.}\BibitemShut {Stop}%
\bibitem [{\citenamefont {Yu}(2025{\natexlab{a}})}]{yu2}%
  \BibitemOpen
  \bibfield  {author} {\bibinfo {author} {\bibfnamefont {Y.}~\bibnamefont
  {Yu}},\ }\href {https://arxiv.org/abs/2504.09845} {\bibfield  {journal}
  {\bibinfo  {journal} {arXiv:2504.09845}\ } (\bibinfo {year}
  {2025}{\natexlab{a}})}\BibitemShut {NoStop}%
\bibitem [{\citenamefont {V.~B.~Berestetskii}\ and\ \citenamefont
  {Pitaevskii}(2008)}]{berestetskiiQEDbook}%
  \BibitemOpen
  \bibfield  {author} {\bibinfo {author} {\bibfnamefont {E.~M.~L.}\
  \bibnamefont {V.~B.~Berestetskii}}\ and\ \bibinfo {author} {\bibfnamefont
  {L.~P.}\ \bibnamefont {Pitaevskii}},\ }\href@noop {} {\emph {\bibinfo {title}
  {Quantum Electrodynamics}}},\ \bibinfo {edition} {2nd}\ ed.\ (\bibinfo
  {publisher} {Elsevier (Singapore) Pte Ltd.},\ \bibinfo {address}
  {Singapore},\ \bibinfo {year} {2008})\BibitemShut {NoStop}%
\bibitem [{\citenamefont {White}(1981)}]{WhiteAtomgasTlBa}%
  \BibitemOpen
  \bibfield  {author} {\bibinfo {author} {\bibfnamefont {J.~C.}\ \bibnamefont
  {White}},\ }\href {\doibase 10.1364/OL.6.000242} {\bibfield  {journal}
  {\bibinfo  {journal} {Optics Letters}\ }\textbf {\bibinfo {volume} {6}},\
  \bibinfo {pages} {242} (\bibinfo {year} {1981})}\BibitemShut {NoStop}%
\bibitem [{\citenamefont {Pedrozo-Pe\~nafiel}\ \emph
  {et~al.}(2012)\citenamefont {Pedrozo-Pe\~nafiel}, \citenamefont {Paiva},
  \citenamefont {Vivanco}, \citenamefont {Bagnato},\ and\ \citenamefont
  {Farias}}]{NaAtoms}%
  \BibitemOpen
  \bibfield  {author} {\bibinfo {author} {\bibfnamefont {E.}~\bibnamefont
  {Pedrozo-Pe\~nafiel}}, \bibinfo {author} {\bibfnamefont {R.~R.}\ \bibnamefont
  {Paiva}}, \bibinfo {author} {\bibfnamefont {F.~J.}\ \bibnamefont {Vivanco}},
  \bibinfo {author} {\bibfnamefont {V.~S.}\ \bibnamefont {Bagnato}}, \ and\
  \bibinfo {author} {\bibfnamefont {K.~M.}\ \bibnamefont {Farias}},\ }\href
  {\doibase 10.1103/PhysRevLett.108.253004} {\bibfield  {journal} {\bibinfo
  {journal} {Phys. Rev. Lett.}\ }\textbf {\bibinfo {volume} {108}},\ \bibinfo
  {pages} {253004} (\bibinfo {year} {2012})}\BibitemShut {NoStop}%
\bibitem [{\citenamefont {Leite}\ and\ \citenamefont
  {Araujo}(1980)}]{rios1980lineshape}%
  \BibitemOpen
  \bibfield  {author} {\bibinfo {author} {\bibfnamefont {J.~R.}\ \bibnamefont
  {Leite}}\ and\ \bibinfo {author} {\bibfnamefont {C.~B.~D.}\ \bibnamefont
  {Araujo}},\ }\href {\doibase 10.1016/0009-2614(80)85514-3} {\bibfield
  {journal} {\bibinfo  {journal} {Chemical Physics Letters}\ }\textbf {\bibinfo
  {volume} {73}},\ \bibinfo {pages} {71} (\bibinfo {year} {1980})}\BibitemShut
  {NoStop}%
\bibitem [{\citenamefont {Andrews}\ and\ \citenamefont
  {Harlow}(1983)}]{andrews1983cooperative}%
  \BibitemOpen
  \bibfield  {author} {\bibinfo {author} {\bibfnamefont {D.~L.}\ \bibnamefont
  {Andrews}}\ and\ \bibinfo {author} {\bibfnamefont {M.}~\bibnamefont
  {Harlow}},\ }\href@noop {} {\bibfield  {journal} {\bibinfo  {journal} {The
  Journal of Chemical Physics}\ }\textbf {\bibinfo {volume} {78}},\ \bibinfo
  {pages} {1088} (\bibinfo {year} {1983})}\BibitemShut {NoStop}%
\bibitem [{\citenamefont {Nayfeh}\ and\ \citenamefont
  {Hillard}(1984)}]{Nayfeh1984}%
  \BibitemOpen
  \bibfield  {author} {\bibinfo {author} {\bibfnamefont {M.~H.}\ \bibnamefont
  {Nayfeh}}\ and\ \bibinfo {author} {\bibfnamefont {G.~B.}\ \bibnamefont
  {Hillard}},\ }\href {\doibase 10.1103/PhysRevA.29.1907} {\bibfield  {journal}
  {\bibinfo  {journal} {Physical Review A}\ }\textbf {\bibinfo {volume} {29}},\
  \bibinfo {pages} {1907} (\bibinfo {year} {1984})}\BibitemShut {NoStop}%
\bibitem [{\citenamefont {Kim}\ and\ \citenamefont {Agarwal}(1998)}]{Kim1998}%
  \BibitemOpen
  \bibfield  {author} {\bibinfo {author} {\bibfnamefont {M.~S.}\ \bibnamefont
  {Kim}}\ and\ \bibinfo {author} {\bibfnamefont {G.~S.}\ \bibnamefont
  {Agarwal}},\ }\href {\doibase 10.1103/PhysRevA.57.3059} {\bibfield  {journal}
  {\bibinfo  {journal} {Physical Review A}\ }\textbf {\bibinfo {volume} {57}},\
  \bibinfo {pages} {3059} (\bibinfo {year} {1998})}\BibitemShut {NoStop}%
\bibitem [{\citenamefont {Muthukrishnan}\ \emph {et~al.}(2004)\citenamefont
  {Muthukrishnan}, \citenamefont {Agarwal},\ and\ \citenamefont
  {Scully}}]{Muthukrishnan2004}%
  \BibitemOpen
  \bibfield  {author} {\bibinfo {author} {\bibfnamefont {A.}~\bibnamefont
  {Muthukrishnan}}, \bibinfo {author} {\bibfnamefont {G.~S.}\ \bibnamefont
  {Agarwal}}, \ and\ \bibinfo {author} {\bibfnamefont {M.~O.}\ \bibnamefont
  {Scully}},\ }\href {\doibase 10.1103/PhysRevLett.93.093002} {\bibfield
  {journal} {\bibinfo  {journal} {Physical Review Letters}\ }\textbf {\bibinfo
  {volume} {93}},\ \bibinfo {pages} {093002} (\bibinfo {year}
  {2004})}\BibitemShut {NoStop}%
\bibitem [{\citenamefont {Zheng}\ \emph {et~al.}(2013)\citenamefont {Zheng},
  \citenamefont {Saldanha}, \citenamefont {Leite},\ and\ \citenamefont
  {Fabre}}]{Zheng2013}%
  \BibitemOpen
  \bibfield  {author} {\bibinfo {author} {\bibfnamefont {Z.}~\bibnamefont
  {Zheng}}, \bibinfo {author} {\bibfnamefont {P.~L.}\ \bibnamefont {Saldanha}},
  \bibinfo {author} {\bibfnamefont {J.~R.~R.}\ \bibnamefont {Leite}}, \ and\
  \bibinfo {author} {\bibfnamefont {C.}~\bibnamefont {Fabre}},\ }\href
  {\doibase 10.1103/PhysRevA.88.033822} {\bibfield  {journal} {\bibinfo
  {journal} {Physical Review A}\ }\textbf {\bibinfo {volume} {88}},\ \bibinfo
  {pages} {033822} (\bibinfo {year} {2013})}\BibitemShut {NoStop}%
\bibitem [{\citenamefont {Hettich}\ \emph {et~al.}(2002)\citenamefont
  {Hettich}, \citenamefont {Schmitt}, \citenamefont {Zitzmann}, \citenamefont
  {K{\"u}hn}, \citenamefont {Gerhardt},\ and\ \citenamefont
  {Sandoghdar}}]{twoMolecules}%
  \BibitemOpen
  \bibfield  {author} {\bibinfo {author} {\bibfnamefont {C.}~\bibnamefont
  {Hettich}}, \bibinfo {author} {\bibfnamefont {C.}~\bibnamefont {Schmitt}},
  \bibinfo {author} {\bibfnamefont {J.}~\bibnamefont {Zitzmann}}, \bibinfo
  {author} {\bibfnamefont {S.}~\bibnamefont {K{\"u}hn}}, \bibinfo {author}
  {\bibfnamefont {I.}~\bibnamefont {Gerhardt}}, \ and\ \bibinfo {author}
  {\bibfnamefont {V.}~\bibnamefont {Sandoghdar}},\ }\href {\doibase
  10.1126/science.1075606} {\bibfield  {journal} {\bibinfo  {journal}
  {Science}\ }\textbf {\bibinfo {volume} {298}},\ \bibinfo {pages} {385}
  (\bibinfo {year} {2002})}\BibitemShut {NoStop}%
\bibitem [{\citenamefont {Yu}(2025{\natexlab{b}})}]{yu1}%
  \BibitemOpen
  \bibfield  {author} {\bibinfo {author} {\bibfnamefont {Y.}~\bibnamefont
  {Yu}},\ }\href {https://arxiv.org/abs/2504.05773} {\bibfield  {journal}
  {\bibinfo  {journal} {arXiv:2504.05773}\ } (\bibinfo {year}
  {2025}{\natexlab{b}})}\BibitemShut {NoStop}%
\bibitem [{\citenamefont {Macfarlane}\ \emph {et~al.}(1992)\citenamefont
  {Macfarlane}, \citenamefont {Cassanho},\ and\ \citenamefont
  {Meltzer}}]{macfarlane1992inhomogeneous}%
  \BibitemOpen
  \bibfield  {author} {\bibinfo {author} {\bibfnamefont {R.}~\bibnamefont
  {Macfarlane}}, \bibinfo {author} {\bibfnamefont {A.}~\bibnamefont
  {Cassanho}}, \ and\ \bibinfo {author} {\bibfnamefont {R.}~\bibnamefont
  {Meltzer}},\ }\href@noop {} {\bibfield  {journal} {\bibinfo  {journal}
  {Physical review letters}\ }\textbf {\bibinfo {volume} {69}},\ \bibinfo
  {pages} {542} (\bibinfo {year} {1992})}\BibitemShut {NoStop}%
\bibitem [{\citenamefont {Macfarlane}\ \emph {et~al.}(1998)\citenamefont
  {Macfarlane}, \citenamefont {Meltzer},\ and\ \citenamefont
  {Malkin}}]{macfarlane1998optical}%
  \BibitemOpen
  \bibfield  {author} {\bibinfo {author} {\bibfnamefont {R.}~\bibnamefont
  {Macfarlane}}, \bibinfo {author} {\bibfnamefont {R.}~\bibnamefont {Meltzer}},
  \ and\ \bibinfo {author} {\bibfnamefont {B.}~\bibnamefont {Malkin}},\
  }\href@noop {} {\bibfield  {journal} {\bibinfo  {journal} {Physical Review
  B}\ }\textbf {\bibinfo {volume} {58}},\ \bibinfo {pages} {5692} (\bibinfo
  {year} {1998})}\BibitemShut {NoStop}%
\bibitem [{\citenamefont {Chukalina}\ \emph {et~al.}(2000)\citenamefont
  {Chukalina}, \citenamefont {Popova}, \citenamefont {Korableva},\ and\
  \citenamefont {Abdulsabirov}}]{chukalina2000fine}%
  \BibitemOpen
  \bibfield  {author} {\bibinfo {author} {\bibfnamefont {E.}~\bibnamefont
  {Chukalina}}, \bibinfo {author} {\bibfnamefont {M.}~\bibnamefont {Popova}},
  \bibinfo {author} {\bibfnamefont {S.}~\bibnamefont {Korableva}}, \ and\
  \bibinfo {author} {\bibfnamefont {R.~Y.}\ \bibnamefont {Abdulsabirov}},\
  }\href@noop {} {\bibfield  {journal} {\bibinfo  {journal} {Physics Letters
  A}\ }\textbf {\bibinfo {volume} {269}},\ \bibinfo {pages} {348} (\bibinfo
  {year} {2000})}\BibitemShut {NoStop}%
\bibitem [{\citenamefont {Liu}\ and\ \citenamefont
  {Jacquier}(2006)}]{liu2006spectroscopic}%
  \BibitemOpen
  \bibfield  {author} {\bibinfo {author} {\bibfnamefont {G.}~\bibnamefont
  {Liu}}\ and\ \bibinfo {author} {\bibfnamefont {B.}~\bibnamefont {Jacquier}},\
  }\href@noop {} {\emph {\bibinfo {title} {Spectroscopic properties of rare
  earths in optical materials}}},\ Vol.~\bibinfo {volume} {83}\ (\bibinfo
  {publisher} {Springer Science \& Business Media},\ \bibinfo {year}
  {2006})\BibitemShut {NoStop}%
\bibitem [{\citenamefont {Weinberg}(1965)}]{Weinberg1965}%
  \BibitemOpen
  \bibfield  {author} {\bibinfo {author} {\bibfnamefont {S.}~\bibnamefont
  {Weinberg}},\ }\href {\doibase 10.1103/PhysRev.140.B516} {\bibfield
  {journal} {\bibinfo  {journal} {Physical Review}\ }\textbf {\bibinfo {volume}
  {140}},\ \bibinfo {pages} {B516} (\bibinfo {year} {1965})}\BibitemShut
  {NoStop}%
\bibitem [{\citenamefont {García-Cely}\ and\ \citenamefont
  {Ringwald}(2024)}]{Garcia-Cely2024}%
  \BibitemOpen
  \bibfield  {author} {\bibinfo {author} {\bibfnamefont {C.}~\bibnamefont
  {García-Cely}}\ and\ \bibinfo {author} {\bibfnamefont {A.}~\bibnamefont
  {Ringwald}},\ }\href {https://arxiv.org/abs/2407.18297} {\bibfield  {journal}
  {\bibinfo  {journal} {arXiv:2407.18297}\ } (\bibinfo {year}
  {2024})}\BibitemShut {NoStop}%
\end{thebibliography}%


\end{document}